\documentclass[12pt]{article}
\newcommand {\citeAY} [1] {\citeNP {#1}}%
\newcommand {\citeAPY}[1] {\citeN  {#1}}%
\usepackage{xchicago}
\usepackage{graphicx}
\usepackage{a4,epsfig}
\renewcommand {\showoriginalref}[1]{}
\renewcommand {\showCODEN}[1]{}
\renewcommand {\showISSN}[1]{}
\renewcommand {\showMR}[3]{}

\newcommand\eq[1] {(\ref{#1})}

\newcommand\fig[1] {\ref{fig:#1}}

\newcommand\labfig[1] {\label{fig:#1}}

\newcommand{\bfm}[1]{\mbox{\boldmath ${#1}$}}
\newcommand{\nonum}{\nonumber \\}
\newcommand{\beqa}{\begin{eqnarray}}
\newcommand{\eeqa}[1]{\label{#1}\end{eqnarray}}
\newcommand{\beq}{\begin{equation}}
\newcommand{\eeq}[1]{\label{#1}\end{equation}}
\newcommand{\Grad}{\nabla}
\newcommand{\Div}{\nabla \cdot}
\newcommand{\Curl}{\nabla \times}

\newcommand{\Tr}{\mathop{\rm Tr}\nolimits}

\newcommand{\lang}{\langle}
\newcommand{\rang}{\rangle}
\newcommand{\Md}{\partial}

\newcommand{\Ga}{\alpha}
\newcommand{\Gb}{\beta}
\newcommand{\Gd}{\delta}
\newcommand{\Ge}{\epsilon}

\newcommand{\Gg}{\gamma}
\newcommand{\Gc}{\chi}

\newcommand{\Gk}{\kappa}

\newcommand{\Gl}{\lambda}
\newcommand{\Gn}{\eta}
\newcommand{\Gm}{\mu}

\newcommand{\Gs}{\sigma}

\newcommand{\GO}{\Omega}

\newcommand{\BGe}{\bfm\epsilon}

\newcommand{\BGn}{\bfm\eta}

\newcommand{\BGt}{\bfm\theta}

\newcommand{\BGs}{\bfm\sigma}

\newcommand{\BGj}{\bfm\tau}

\newcommand{\BGz}{\bfm\zeta}

\newcommand{\CC}{{\cal C}}

\newcommand{\BCA}{{\bfm{\cal A}}}

\newcommand{\BCC}{{\bfm{\cal C}}}

\def\Bb{{\bf b}}

\def\Be{{\bf e}}
\def\Bf{{\bf f}}

\def\Bj{{\bf j}}
\def\Bk{{\bf k}}

\def\Bm{{\bf m}}
\def\Bn{{\bf n}}

\def\Bu{{\bf u}}
\def\Bv{{\bf v}}

\def\Bx{{\bf x}}

\def\BA{{\bf A}}
\def\BB{{\bf B}}

\def\BI{{\bf I}}

\def\BR{{\bf R}}

\def \ba {\begin{array}}
\def \ea {\end{array}}
\def \refe #1.{(\ref{#1})}
\def \reff #1.{figure~\ref{#1}}
\def \refs #1.{section~\ref{#1}}
\def \refss #1.{subsection~\ref{#1}}
\def \refD #1.{Definition~\ref{#1}}
\def \refT #1.{Theorem~\ref{#1}}
\def \refL #1.{Lemma~\ref{#1}}
\def \refC #1.{Corollary~\ref{#1}}
\def \refP #1.{Proposition~\ref{#1}}
\def \refR #1.{Remark~\ref{#1}}
\def \refE #1.{Example~\ref{#1}}
\def \refN #1.{Notation~\ref{#1}}

\usepackage{graphics}
\usepackage{amssymb}
\begin{document}
\vspace{-1in}
\title{Universal bounds on the electrical and elastic response of two-phase bodies and their application to bounding the volume fraction from boundary measurements}
\author{}
\author{Graeme W. Milton\\
\small{Department of Mathematics, University of Utah, Salt Lake City UT 84112, USA}\\
\small{email: milton@math.utah.edu,~~telephone: 1(801)581-6495,~~fax: 1(801)581-4148}}

\date{}
\maketitle
\begin{abstract}
Universal bounds on the electrical and elastic response of two-phase (and multiphase) ellipsoidal or parallelopipedic bodies have been obtained by 
Nemat-Nasser and Hori. Here we show how their bounds can be improved and extended to bodies of arbitrary shape. Although our analysis is for two-phase bodies 
with isotropic phases it can easily be extended to multiphase bodies with anisotropic constituents. Our two-phase bounds can be used in an inverse fashion
to bound the volume fractions occupied by the phases, and for electrical conductivity reduce to those of Capdeboscq and Vogelius 
when the volume fraction is asymptotically small. Other volume fraction bounds derived here
utilize information obtained from thermal, magnetic, dielectric or elastic responses. 
One bound on the volume fraction can be obtained by simply immersing the body in a water filled cylinder with a 
piston at one end and measuring the change in water pressure when the piston is displaced by a known small amount. 
This bound may be particularly effective for estimating the volume of cavities in a body.
We also obtain new bounds utilizing
just one pair of (voltage, flux) electrical measurements at the boundary of the body.

\end{abstract}
\vskip2mm


\noindent 
\vskip2mm
\section{Introduction}
\setcounter{equation}{0}
\citeAPY{Berryman:1990:VCE} found that classical variational principles could be used to obtain information about the
conductivity inside a body from electrical measurements on the exterior. In this paper our main focus is on using classical
variational principles and known bounds on the response of periodic composites to bound
the volume fraction of one phase in a two-phase body $\GO$ from measurements on the exterior of the body. Of course if one
knows the mass densities of the two phases, the easiest way to do this is just to weigh the body. However this 
may not always be practical, or the densities of the two phases may be very close. 

Two types of boundary conditions are most natural: what we call special Dirichlet conditions where affine Dirichlet
conditions are imposed on the boundary of $\GO$ (which would render the field inside $\GO$ uniform if the body were
homogeneous) or what we call special Neumann conditions where Neumann conditions are imposed that would render the field inside $\GO$
uniform if the body were homogeneous. Bounds on the electrical and elastic response of the body to these special boundary conditions were obtained 
by Nemat-Nasser and Hori (\citeyearNP{Nemat-Nasser:1993:MOP}, \citeyearNP{Nemat-Nasser:1995:UBO}), and were extended to piezoelectricity
by \citeAPY{Hori:1998:UBE}. They called these bounds universal because they did not depend on any assumption about the microgeometry in the body.
They obtained both elementary universal bounds based on the classical variational principles, and reviewed below in section 3, and 
universal bounds based on the Hashin-Shtrikman variational principles (Hashin and Shtrikman \citeyearNP{Hashin:1962:VAT}, \citeyearNP{Hashin:1963:VAT}).
The latter bounds were obtained under the assumption that $\GO$ is either an ellipsoid or a parallelopiped, but we will see here that they
can easily be improved and generalized to bodies $\GO$ of arbitrary shape. The key is to consider an assemblage of copies of $\GO$ packed to
fill all space, and then to use the bounds of \citeAPY{Huet:1990:AVC} which relate the effective tensor of this composite to the
responses of $\GO$ under the special boundary conditions. Then existing bounds on the effective tensor [as surveyed in the books
of \citeAPY{Nemat-Nasser:1993:MOP}, \citeAPY{Cherkaev:2000:VMS}, \citeAPY{Allaire:2002:SOH}, \citeAPY{Milton:2002:TC}, and \citeAPY{Tartar:2009:GTH}]
can be directly applied to bound the responses of $\GO$ under special boundary conditions (see sections 5,6,7, and 8).
Since these bounds involve the volume fractions of the phases
(and the moduli of the phases), they can be used in an inverse fashion to bound the volume fraction. As shown by \citeAPY{Kang:2011:SBV}
the volume fraction bounds thus obtained 
for electrical conductivity generalize those obtained by Capdeboscq and Vogelius (\citeyearNP{Capdeboscq:2003:OAE}, \citeyearNP{Capdeboscq:2004:RSR})
for the important case when the volume fraction is asymptotically small.

 Given the close connection between bounds on effective tensors
and bounds on the responses of $\GO$ under special boundary condition, a natural question to ask is whether methods that have been used
to derive bounds on effective tensors could be directly used to derive bounds on the response of $\GO$ under more general boundary conditions.
One such method is the Hashin-Strikman (\citeyearNP{Hashin:1962:VAT}, \citeyearNP{Hashin:1963:VAT}) variational method and this lead
Nemat-Nasser and Hori to their bounds for ellipsoidal or parallelopipedic $\GO$. Another particularly successful method is the translation method 
(\citeAY{Tartar:1979:ECH}; Lurie and Cherkaev \citeyearNP{Lurie:1982:AEC}, \citeyearNP{Lurie:1984:EEC}; \citeAY{Murat:1985:CVH}; \citeAY{Tartar:1985:EFC}; \citeAY{Milton:1990:CSP})
and indeed as shown in a companion paper (\citeAY{Kang:2011:SBV}) this method yields upper and lower bounds on the volume fraction in a
two-phase body with general boundary conditions for two-dimensional conductivity without making any assumption on the shape of $\GO$.
For special boundary conditions the bounds thus derived reduce to the ones derived here. 

We also provide (in section 4) some new conductivity bounds which just involve the results of just one (flux, voltage) pair
measured at the boundary of $\GO$, and which improve upon the elementary bounds of \citeAPY{Nemat-Nasser:1993:MOP}.
Again these new bounds can be used in an inverse fashion to bound the volume fraction. Other volume fraction
bounds using one measurement were derived by  \citeAPY{Kang:1997:ICP}
\citeAPY{Ikehata:1998:SEI}, \citeAPY{Alessandrini:1998:ICP}, \citeAPY{Alessandrini:2000:OSE}, and \citeAPY{Alessandrini:2002:DCE}.
These other bounds involve constants which are not easy to determine, making it difficult to make a general comparison with
our new bounds. 

The various bounds on the volume fraction we have derived are too numerous to summarize in this introduction. 
However we want to draw attention to the bounds \eq{3.12} and \eq{3.21}  which are the natural extension of the
famous \citeAPY{Hashin:1962:VAT} conductivity bounds to this problem. Also of particular note is the bound \eq{5.18ag},
which is one natural generalization of the bulk modulus bounds of \citeAPY{Hashin:1963:VAT} and \citeAPY{Hill:1963:EPR},
and implies that a bound on the volume fraction can be obtained by simply immersing the body in a water filled 
cylinder with a piston at one end and measuring the change in water pressure when the piston is displaced by a known
small amount.

\section{The conductivity response tensors with special Dirichlet and special Neumann boundary conditions}
\setcounter{equation}{0}

In electrical impedance tomography in a body $\GO$ containing two isotropic components with (positive, scalar) 
conductivities $\Gs_1$ and $\Gs_2$ the potential $V$ satisfies 
\beq \Div\Gs\Grad V=0, \quad {\rm where}\quad \Gs(\Bx)=\Gc(\Bx)\Gs_1+(1-\Gc(\Bx))\Gs_2,
\eeq{1.1}
and $\Gc(\Bx)$ is the indicator function of component $1$, taking the value $1$ in component and $0$ in component
$2$. Equivalently, in terms of the current field $\Bj(\Bx)$ and electric field $\Be(\Bx)$ we have
\beq \Div\Bj=0,\quad\Bj=\Gs\Be,\quad \Be=-\Grad V. \eeq{1.2} 
Let us assume the components have been labeled so that $\Gs_1\geq\Gs_2$. We are given a set of Cauchy data, i.e. 
measurements of pairs $(V_0,q)$, where $V_0(\Bx)$ and $q(\Bx)$ are the boundary values of the voltage $V(\Bx)$ and
and flux $q(\Bx)=-\Bn\cdot\Bj(\Bx)$ at the boundary $\partial\GO$ of $\GO$, in which $\Bn(\Bx)$ is the outwards 
normal to the boundary. From this boundary information we can immediately determine, using integration by parts, 
volume averages such as
\beqa \lang \Be\cdot\Bj\rang &= & \frac{1}{|\GO|}\int_{\partial\GO}-V_0(\Bj\cdot\Bn)= \frac{1}{|\GO|}\int_{\partial\GO}V_0q, \nonum
      \lang \Be\rang & = & \frac{1}{|\GO|}\int_{\partial\GO}-V_0\Bn, \nonum
      \lang \Bj\rang & = & \frac{1}{|\GO|}\int_{\partial\GO}-\Bx q,
\eeqa{1.3}
where the angular brackets denote the volume average, i.e.
\beq \lang g \rang=\frac{1}{|\GO|}\int_{\GO} g, \eeq{1.4}
for any quantity $g(\Bx)$. From such averages our objective is to bound the
volume fraction $f_1=\lang \Gc\rang$ of component 1 (and hence also the volume fraction $f_2=1-f_1$ of component
2).

To obtain good estimates of the volume fraction $f_1$ it makes physical sense to use
measurements where the fields $\Be(\Bx)$ and $\Bj(\Bx)$ probe well into the interior of $\GO$. In this
connection two sets of measurements are most natural. We could apply special Dirichlet boundary conditions
\beq V_0=-\Be_0\cdot\Bx, \eeq{1.5}
and measure $\Bj_0=\lang\Bj\rang$. Here, according to \eq{1.3}, $\Be_0$ equals $\lang\Be\rang$. Since
$\Bj_0$ is linearly related to $\Be_0$ we can write
\beq \Bj_0=\BGs^D\Be_0, \eeq{1.6}  
which defines the conductivity tensor $\BGs^D$ ($D$ for Dirichlet). To determine $\BGs^D$ 
in dimension $d=2,3$ it of course suffices to measure $\Bj_0$ for $d$ linearly independent values of $\Be_0$. 
Alternatively we could apply the special  Neumann boundary conditions 
\beq q=\Bj_0\cdot\Bn, \eeq{1.7}
and measure $\Be_0=\lang\Be\rang$. Again according to \eq{1.3}, $\Bj_0=\lang\Bj\rang$. The linear relation
between $\Be_0$ and $\Bj_0$,
\beq \Be_0=(\BGs^N)^{-1}\Bj_0 \eeq{1.8}
defines the resistivity tensor $(\BGs^N)^{-1}$ and hence the conductivity tensor $\BGs^N$ ($N$ for Neumann):
we will see later that $(\BGs^N)^{-1}$ is invertible. To determine $\BGs^N$ it suffices to measure $\Be_0$ 
for $d$ linearly independent values of $\Bj_0$.  With either of these two sorts of boundary conditions (but 
not in general) \citeAPY{Hill:1963:EPR} has shown that
\beq \lang \Be\cdot\Bj\rang=\lang\Be\rang\cdot\lang\Bj\rang, \eeq{1.9}
as follows by substituting \eq{1.5} or \eq{1.7} in the first of equations \eq{1.3}.
Using this relationship, and its obvious generalizations, it is easy to check that both $\BGs^D$ and 
$\BGs^N$ are self-adjoint. Thus if $\Be'(\Bx)$ and $\Bj'(\Bx)$ denote the electric and current fields
associated with the boundary conditions \eq{1.5}, with $\Be_0$ replaced by $\Be_0'$, while keeping
the same conductivity $\Gs(\Bx)$ then
\beq \Be_0'\cdot\BGs^D\Be_0=\lang\Be'\cdot\Bj\rang=\lang\Be'\Gs\Be\rang
=\lang\Be\cdot\Bj'\rang=\Be_0\cdot\BGs^D\Be_0',
\eeq{1.10}
which implies $\BGs^D$ is self-adjoint. By similar argument $\BGs^N$ is self-adjoint.

\section{Known elementary bounds}
\setcounter{equation}{0}

This section reviews the elementary bounds on  $\BGs^D$ and $\BGs^N$ obtained by
\citeAPY{Nemat-Nasser:1993:MOP} and by Willis
in a 1989 private communication to Nemat-Nasser and Hori. Their implications for bounding the volume fraction
will be studied. We will make use of two classical variational principles: the Dirichlet variational principle that
\beq \min_{\matrix{{\underline{\Be}}(\Bx)=-\Grad{\underline{V}}(\Bx) 
\cr {\underline{V}}(\Bx)=V_0(\Bx)~{\rm on}~\partial\GO}}\int_{\GO}\underline{\Be}\cdot\Gs\underline{\Be}
=\int_{\partial\GO}-V_0q,
\eeq{2.1}
which is attained when $\underline{V}(\Bx)=V(\Bx)$, and the Neumann variational principle that
\beq \min_{\matrix{{\underline{\Bj}}(\Bx)\cr \Div{\underline{\Bj}}(\Bx)=0 \cr 
     \Bn\cdot{\underline{\Bj}}(\Bx)=-q(\Bx)~{\rm on}~\partial\GO}}\int_{\GO}\underline{\Bj}\cdot\Gs^{-1}\underline{\Bj}
=\int_{\partial\GO}V_0q,
\eeq{2.2}
which is attained when $\underline{j}(\Bx)=\Bj(\Bx)$. With the special Dirichlet boundary conditions
\eq{1.5} the Dirichlet variational principle implies 
\beq \min_{\matrix{{\underline{\Be}}(\Bx)=-\Grad{\underline{V}}(\Bx) 
\cr {\underline{V}}(\Bx)=-\Be_0\cdot\Bx~{\rm on}~\partial\GO}}\lang\underline{\Be}\cdot\Gs\underline{\Be}\rang
=\Be_0\cdot\BGs^D\Be_0.
\eeq{2.3}
Taking a trial potential $\underline{V}=-\Be_0\cdot\Bx$ produces the elementary upper bound on $\Be_0\cdot\BGs^D\Be_0$
\beq \Be_0\cdot\BGs^D\Be_0\leq \lang\Gs\rang\Be_0\cdot\Be_0,
\eeq{2.4}
given by \citeAPY{Nemat-Nasser:1993:MOP}.
To obtain a lower bound observe, following a standard argument, that the left hand side of \eq{2.3} is surely
decreased if we take the minimum over a larger class of trial fields. Since the constraints on 
${\underline{\Be}}(\Bx)$ imply $\lang{\underline{\Be}}\rang=\Be_0$ let us replace them by this weaker constraint to 
obtain the inequality
\beq \Be_0\cdot\BGs^D\Be_0\geq \min_{\matrix{{\underline{\Be}}(\Bx)\cr\lang{\underline{\Be}}\rang=\Be_0}}
\lang\underline{\Be}\cdot\Gs\underline{\Be}\rang,
\eeq{2.5}
where the minimum is now over fields ${\underline{\Be}}$ which are not necessarily curl-free. Using Lagrange
multipliers one finds that the minimum is attained when ${\underline{\Be}}=\Gs^{-1}\lang\Gs^{-1}\rang^{-1}\Be_0$
and so we obtain the lower bound 
\beq \Be_0\cdot\BGs^D\Be_0\geq \lang\Gs^{-1}\rang^{-1}\Be_0\cdot\Be_0
\eeq{2.6}
of \citeAPY{Nemat-Nasser:1993:MOP}.
Taken together, \eq{2.4} and \eq{2.6} imply the lower and upper bounds
\beq \left(\frac{\Be_0\cdot\BGs^D\Be_0}{\Be_0\cdot\Be_0}-\Gs_2\right)/(\Gs_1-\Gs_2)
\leq f_1 \leq \left(\Gs_2^{-1}-\frac{\Be_0\cdot\Be_0}{\Be_0\cdot\BGs^D\Be_0}\right)/(\Gs_2^{-1}-\Gs_1^{-1})
\eeq{2.7}
on the volume fraction $f_1$. These bounds give useful information even if we only know $\Bj_0=\BGs^D\Be_0$
for only one value of $\Be_0$, i.e. if we only take one measurement. These bounds \eq{2.7} are sharp in 
the sense that the lower bound is approached artitrarily closely if $\GO$ is filled with a periodic laminate
of components 1 and 2, oriented with the normal to the layers orthogonal to $\Be_0$ and we let the period
length go to zero, while the upper bound is approached artitrarily closely for the same geometry, but 
oriented with the normal to the layers parallel to $\Be_0$. If the full tensor $\BGs^D$ is known,
from $d=2,3$ measurements of pairs $(\Be_0,\Bj_0)$  
then we can take the intersection of the bounds \eq{2.7} as $\Be_0$ is varied, and so obtain 
\beq (\Gl^D_+-\Gs_2)/(\Gs_1-\Gs_2)
\leq f_1 \leq (1/\Gs_2-1/\Gl^D_-)/(\Gs_2^{-1}-\Gs_1^{-1}),
\eeq{2.8}
where $\Gl^D_+$ and $\Gl^D_-$ are the maximum and minimum eigenvalues of $\BGs^D$. However
we will see in the next section that an additional and typically sharper upper bound on $f_1$ can be obtained.

With the special Neumann boundary conditions \eq{1.7} the variational principle \eq{2.2} implies 
\beq \min_{\matrix{{\underline{\Bj}}(\Bx)\cr \Div{\underline{\Bj}}(\Bx)=0 \cr 
     \Bn\cdot{\underline{\Bj}}(\Bx)=\Bn\cdot\Bj_0~{\rm on}~\partial\GO}}
\lang\underline{\Bj}\cdot\Gs^{-1}\underline{\Bj}\rang=\Bj_0\cdot(\BGs^N)^{-1}\Bj_0.
\eeq{2.9}
By taking a constant trial field ${\underline{\Bj}}(\Bx)=\Bj_0$ or alternatively by taking the minimum over the larger
class of trial fields satisfying only $\lang{\underline{\Bj}}\rang=\Bj_0$ we 
obtain the bounds
\beq \lang\Gs\rang^{-1}\Bj_0\cdot\Bj_0\leq\Bj_0\cdot(\BGs^N)^{-1}\Bj_0\leq \lang\Gs^{-1}\rang\Bj_0\cdot\Bj_0
\eeq{2.10}
of \citeAPY{Nemat-Nasser:1993:MOP} which imply
\beq \left(\frac{\Bj_0\cdot\Bj_0}{\Bj_0\cdot(\BGs^N)^{-1}\Bj_0}-\Gs_2\right)/(\Gs_1-\Gs_2)
\leq f_1 \leq \left(\Gs_2^{-1}-\frac{\Bj_0\cdot(\BGs^N)^{-1}\Bj_0}{\Bj_0\cdot\Bj_0}\right)/(\Gs_2^{-1}-\Gs_1^{-1}).
\eeq{2.11}
These bounds are applicable even if we only know $\Be_0=(\BGs^N)^{-1}\Bj_0$
for only one value of $\Bj_0$. For comparison, with these special Neumann boundary conditions \eq{1.7},
the bounds in Theorem 3.1 of \citeAPY{Kang:1997:ICP} coupled with the improvement in proposition 0
of \citeAPY{Ikehata:1998:SEI},
with $\Gs_1=k>1$, $\Gs_2=1$ and $\Bj_0\cdot\Bj_0=1$, imply
\beq 
\frac{1}{k-1}(1-\Bj_0\cdot(\BGs^N)^{-1}\Bj_0)
\leq f_1 \leq \frac{k}{k-1}(1-\Bj_0\cdot(\BGs^N)^{-1}\Bj_0).
\eeq{2.12}
In this case it is easy to check that the upper bounds in \eq{2.11} and \eq{2.12} coincide while the lower bound
in \eq{2.11} is tighter. The bounds \eq{2.11} are each approached arbitrarily closely if $\GO$ is filled with a periodic laminate
of components 1 and 2, oriented with $\Bj_0$ either parallel or orthogonal to the layers
and we let the period length go to zero.

In summary, \eq{2.4},\eq{2.6} and \eq{2.10} imply the matrix inequalities
\beq  \lang\Gs^{-1}\rang^{-1}\BI\leq\BGs^D\leq \lang\Gs\rang\BI,\quad\quad
 \lang\Gs^{-1}\rang^{-1}\BI\leq\BGs^N\leq \lang\Gs\rang\BI
\eeq{2.13}
of \citeAPY{Nemat-Nasser:1993:MOP}.

For artitrary boundary conditions, i.e. for any $\Be$ and $\Bj$ satisfying \eq{1.2} within $\GO$,
we have the bounds 
\beq \lang\Be\cdot\Bj\rang \geq \Be_0\cdot\BGs^N\Be_0,\quad
\lang\Be\cdot\Bj\rang \geq \Bj_0(\BGs^D)^{-1}\Bj_0,
\eeq{2.14}
where $\Be_0=\lang\Be\rang$ and $\Bj_0=\lang\Bj\rang$,
due to Willis in a 1989 private communication to Nemat-Nasser and Hori, and presented by \citeAPY{Nemat-Nasser:1993:MOP}. In conjunction with \eq{2.13} they imply
the volume fraction bounds,
\beq \left(\frac{\Bj_0\cdot\Bj_0}{\lang\Be\cdot\Bj\rang}-\Gs_2\right)/(\Gs_1-\Gs_2)
\leq f_1 \leq \left(\Gs_2^{-1}-\frac{\Be_0\cdot\Be_0}{\lang\Be\cdot\Bj\rang}\right)/(\Gs_2^{-1}-\Gs_1^{-1}).
\eeq{2.15}

\section{New bounds with one measurement}
\setcounter{equation}{0}

If we have measurements of $\lang\Be\cdot\Bj\rang$ and both vectors $\Be_0$ and $\Bj_0$ for arbitrary boundary 
conditions then the bounds \eq{2.14} and \eq{2.15} can be improved. The classical variational principle
\eq{2.1} implies 
\beq
\min_{\matrix{{\underline{\Be}}(\Bx)=-\Grad{\underline{V}}(\Bx) 
\cr {\underline{V}}(\Bx)=V_0(\Bx)~{\rm on}~\partial\GO \cr
\lang\sigma\underline{\Be}\rang=\Bj_0}}\lang\underline{\Be}\cdot\Gs\underline{\Be}\rang,
=\lang\Be\cdot\Bj\rang
\eeq{2.16}
where we have chosen to add the constraint that $\lang\sigma\underline{\Be}\rang=\Bj_0$ since we know that
without this constraint the minimizer $\underline{\Be}=\Be$ satisfies $\lang\sigma\Be\rang=\Bj_0$. We surely
obtain something lower if take the minimum over the larger class of fields satisfying only
$\lang\underline{\Be}\rang=\Be_0$ and $\lang\sigma\underline{\Be}\rang=\Bj_0$. Thus we obtain the inequality
\beq 
\min_{\matrix{{\underline{\Be}}(\Bx)\cr \lang\underline{\Be}\rang=\Be_0 \cr \lang\sigma\underline{\Be}\rang=\Bj_0}}
\lang\underline{\Be}\cdot\Gs\underline{\Be}\rang\leq\lang\Be\cdot\Bj\rang.
\eeq{2.17}
By introducing two vector valued Lagrange multipliers associated with the two vector valued constraints we find that
the minimum is attained when
\beq \underline{\Be}(\Bx)
=\Be_0+(\lang\Gs^{-1}\rang-\Gs^{-1}(\Bx))(\lang\Gs\rang\lang\Gs^{-1}\rang-1)^{-1}(\Bj_0-\lang\Gs\rang\Be_0).
\eeq{2.18}
Substituting this back in \eq{2.17} gives the bound
\beq (\lang\Gs\rang\lang\Gs^{-1}\rang-1)(\lang\Be\cdot\Bj\rang-\Be_0\cdot\Bj_0)
\geq (\Bj_0-\lang\Gs\rang\Be_0)\cdot(\lang\Gs^{-1}\rang\Bj_0-\Be_0).
\eeq{2.19}
If, with general boundary conditions, we are interested in bounding the volume fraction $f_1$ given measured values of 
$\lang\Be\cdot\Bj\rang$, $\Be_0$ and $\Bj_0$ then the difference between the left hand side and right hand side
of \eq{2.19} is a quadratic in $f_1$ whose two roots give upper and lower bounds on $f_1$. (Unless the roots happen
to be complex, in which case there is no configuration of the two phases within $\GO$ which produce the measured
$\lang\Be\cdot\Bj\rang$, $\Be_0$ and $\Bj_0$, indicating the presence of other phases or indicating
an error in measurements.)

In the particular cases of either special Dirichlet or special Neumann boundary conditions, 
\eq{1.5} or \eq{1.7}, the left hand
side of \eq{2.19} vanishes (see \eq{1.9}) and we obtain the reduced bounds
\beq 0\geq (\Bj_0-\lang\Gs\rang\Be_0)\cdot(\lang\Gs^{-1}\rang\Bj_0-\Be_0),
\eeq{2.20}
which are in fact implied by the matrix inequalities \eq{2.13}. This bound \eq{2.20} is optimal.
For any given fixed $\Be_0$, and fixed volume fraction $f_1$, the vector $\Bj_0$ 
has an endpoint which is constrained by \eq{2.20} to lie within a
sphere (disk in two dimensions) centered at $(\lang\Gs\rang+\lang\Gs^{-1}\rang^{-1})\Be_0/2$.
When $\GO$ is filled with a periodic laminate of the two phases
with interfaces orthogonal to some unit vector $\Bm$, and we let the period length go to zero,
then the endpoint of the vector $\Bj_0$ covers the entire surface of this sphere (disk) as $\Bm$
ranges over all unit vectors. These bounds are the analogs, for arbitrary bodies $\GO$, of 
bounds on possible $(\Be_0,\Bj_0)$ pairs
for composites derived by \citeAPY{Raitum:1983:QES} and \citeAPY{Tartar:1995:RHM}. If we are given $\Be_0$ and $\Bj_0$
and want to bound $f_1$ then we should find the range of $f_1$ where the sphere (or disk)
contains the endpoint of the vector $\Bj_0$. The endpoints of this range are the roots
of the right hand side of \eq{2.20} which is a quadratic function of $f_1$.  

Knowledge of $\Be_0$ and $\Bj_0$ is equivalent to knowledge of $\lang\Be\cdot\Bv\rang$
and $\lang\Bv\cdot\Bj\rang$ for all constant fields $\Bv$. A more general alternative is to use
the information about  
\beqa a_i=\lang \Be\cdot\Bj_i\rang &= & \frac{1}{|\GO|}\int_{\partial\GO}-V_0(\Bj_i\cdot\Bn),
\nonum
 b_k=\lang \Grad V_k\cdot\Bj\rang &= & \frac{1}{|\GO|}\int_{\partial\GO}-V_k(\Bj\cdot\Bn),
\eeqa{2.21}
for a given set of ``comparison flux fields'' $\Bj_i(\Bx)$ satisfying $\Div\Bj_i=0$,
$i=1,2,\ldots n$ and ``comparison potentials'' $V_k(\Bx),~k=1,2,\ldots m$. Suppose, for
example, that we have just one comparison flux field $\Bj_1$. We have the variational
principle
\beq
\min_{\matrix{{\underline{\Be}}(\Bx)=-\Grad{\underline{V}}(\Bx) 
\cr {\underline{V}}(\Bx)=V_0(\Bx)~{\rm on}~\partial\GO \cr
\lang \underline{\Be}\cdot\Bj_1\rang=a_1}}\lang\underline{\Be}\cdot\Gs\underline{\Be}\rang
=\lang\Be\cdot\Bj\rang,
\eeq{2.22}
where we have chosen to add the constraint that $\lang\underline{\Be}\cdot\Bj_1\rang=a_1$ since we know that
without this constraint the minimizer $\underline{\Be}=\Be$ 
satisfies $\lang\Be\cdot\Bj_1\rang=a_1$. This implies the inequality
\beq \lang\Be\cdot\Bj\rang\geq \min_{\matrix{{\underline{\Be}}(\Bx)\cr 
\lang \underline{\Be}\cdot\Bj_1\rang=a_1}}\lang\underline{\Be}\cdot\Gs\underline{\Be}\rang.   
\eeq{2.23}
By introducing a Lagrange multiplier associated with the constraint $\lang\Be\cdot\Bj_1\rang=a_1$
we see the minimum occurs when
\beq \underline{\Be}=a_1\Gs^{-1}\Bj_1/\lang\Bj_1\cdot\Gs^{-1}\Bj_1\rang,
\eeq{2.24}
giving the inequality
\beq \lang\Be\cdot\Bj\rang\geq a_1^2/\lang\Bj_1\cdot\Gs^{-1}\Bj_1\rang. \eeq{2.25a}
This inequality gives information about $\Gs(\Bx)$ through $\lang\Bj_1\cdot\Gs^{-1}\Bj_1\rang$.
If we only want bounds which involve the volume fraction we should choose $\Bj_1(\Bx)$
with
\beq |\Bj_1(\Bx)|=1\quad {\rm for~all~}\Bx\in\GO.
\eeq{2.25} 
There are many divergence free fields $\Bj_1(\Bx)$ which satisfy this constraint. For example
in two dimensions we can take
\beq \Bj_1=({\Md \phi/\Md x_2,-\Md \phi/\Md x_1}),~~
{\rm with}~|\Grad\phi(\Bx)|=1\quad {\rm for~all~}\Bx\in\GO.
\eeq{2.26}
Thus $\phi(\Bx)$ satisfies an Eikonal equation, and we could take $\phi(\Bx)$ to be the shortest
distance between $\Bx$ and a curve outside $\GO$. Once \eq{2.25} is satisfied \eq{2.25a}
implies the volume fraction bound
\beq f_1\leq 
 \left(\Gs_2^{-1}-\frac{a_1^2}{\lang\Be\cdot\Bj\rang}\right)/(\Gs_2^{-1}-\Gs_1^{-1}).
\eeq{2.27}
In the special case when $\Bj_1=\Be_0/|\Be_0|$ this reduces to the upper bound on 
$f_1$ given by \eq{2.15}. 

An important question is whether this new bound is sharp, and if so for what $\Gs(\Bx)$?
The new bound will be sharp when $\Be=\underline{\Be}$ where $\underline{\Be}$ is
given by \eq{2.24}. In that case
\beq \Bj(\Bx)=a_1\Bj_1/\lang\Bj_1\cdot\Gs^{-1}\Bj_1\rang
\eeq{2.28}
has zero divergence because it is proportional to $\Bj_1$. 
Let us impose the Neumann boundary condition
\beq \Bj(\Bx)\cdot\Bn=\Bj_1\cdot\Bn\quad {\rm for~all}~\Bx\in\Md\GO,
\eeq{2.29}
and look for a $\Gs(\Bx)$ so $\Bj(\Bx)=\Bj_1(\Bx)$ and $\Be(\Bx)=\Gs^{-1}\Bj_1(\Bx)$
is curl-free. Now as schematically represented by figure \fig{0}, choose
$\Gs(\Bx)$ to correspond to a finely layered composite with layers
orthogonal to the streamlines of $\Bj_1(\Bx)$, and with phase 1 occupying
a local volume fraction $p(\Bx)$. This composite will support a
current field $\Bj(\Bx)=\Bj_1(\Bx)$ and an electric field $\Be(\Bx)=\Gs^{-1}\Bj_1(\Bx)$
provided 
\beq \Curl\Be_0=0,\quad\Be_0\equiv[\Gs_2^{-1}-p(\Bx)(\Gs_2^{-1}-\Gs_1^{-1})]\Bj_1(\Bx).
\eeq{2.30}
Here $\Be_0(\Bx)$ is the weak limit (local volume average) of $\Be(\Bx)$ as the
layer spacing goes to zero. In two dimensions, given $\Bj_1(\Bx)$ we could look for solutions
for $p(\Bx)$ such that \eq{2.30} is satisfied and $0\leq p(\Bx)\leq 1$ in $\GO$. We expect such
solutions to exist for a wide class of fields $\Bj_1(\Bx)$. This example shows that non-constant
``comparison flux fields'' can lead to sharp bounds on the volume fraction. In three
dimensions we only expect to find a solution of the vector equation \eq{2.30}
for the scalar field $p(\Bx)$ if $\Bj_1(\Bx)$ satisfies some additional conditions.

\begin{figure}
\vspace{2in}
\hspace{1.0in}
{\resizebox{2.0in}{1.0in}
{\includegraphics[0in,0in][6in,3in]{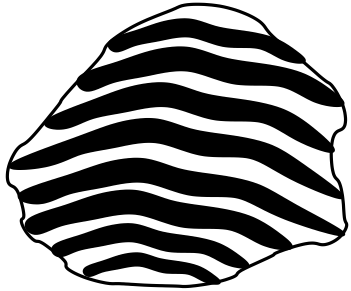}}}
\vspace{0.1in}
\caption{A schematic of the type of layered microstructure achieving the volume fraction bound \eq{2.27}, where the black regions denote one phase, and the
white regions the other phase. The layer widths should be much finer than the size of $\GO$. }
\labfig{0}
\end{figure}

\section{Relationship to bounding effective tensors of composites}
\setcounter{equation}{0}

Consider a periodic composite obtained by taking the unit cell boundaries outside $\GO\equiv\GO_1$
and almost filling the rest of the unit cell by non-intersecting
rescaled and translated copies $\GO_i$, $i=2,\ldots, n$ of
$\GO$, as illustrated in figure \fig{1}. The remainder of the unit cell is filled by phase 2 with
conductivity $\Gs_2$. The unit cell structure is periodically repeated to fill all space.
Let $\Gs_C(\Bx)$ ($C$ for composite) denote this effective conductivity, i.e. in 
the unit cell
\beqa \Gs_C(\Bx) &=& \Gs(\Bx/a_i+\Bb_i)~{\rm in}~\GO_i,~~i=1,2,\ldots, N \nonum
                 &=& \Gs_2~{\rm elsewhere~outside}~\cup_{i}\GO_i,
\eeqa{3.1}
where the scaling constants $a_i$ and translation vectors $\Bb_i$ (with $a_1=1$ and $\Bb_1=0$)
are determined by the size and
position of each copy $\GO_i$, so that $\Bx/a_i+\Bb_i$ is on the boundary of $\GO$ if and only if
$\Bx$ is on the boundary of $\GO_i$. Let $p_n$ denote the volume fraction in the unit cell
occupied by the material with conductivity $\Gs_2$. Let $\BGs^*_n$ denote the (matrix valued) 
effective conductivity of this composite, which in general depends upon the relative positions
of the copies $\GO_i$ within the unit cell.

\begin{figure}
\vspace{2in}
\hspace{1.0in}
{\resizebox{2.0in}{1.0in}
{\includegraphics[0in,0in][6in,3in]{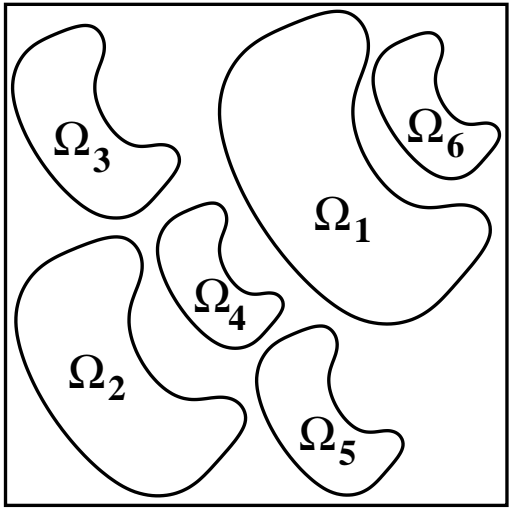}}}
\vspace{0.1in}
\caption{A period cell containing rescaled copies of $\GO$. }
\labfig{1}
\end{figure}

We have the classical variational inequality 
\beq \Be_0\cdot\BGs^*_n\Be_0\leq \lang\underline{\Be}_C\cdot\Gs_C\underline{\Be}_C\rang,
\eeq{3.2}
which holds for any trial electric field $\underline{\Be}_C$ satisfying
\beq \Curl\underline{\Be}_C=0,\quad\underline{\Be}_C~{\rm periodic},\quad
\lang\underline{\Be}_C\rang=\Be_0,
\eeq{3.3}
where now the volume averages are over the entire unit cell, rather than just $\GO$. 
In particular, letting $\Be(\Bx)$ denote the electric field within $\GO$ when the special
Dirichlet boundary conditions \eq{1.5} are applied, we may take in the unit cell
\beqa \underline{\Be}_C(\Bx) &=& \Be(\Bx/a_i+\Bb_i)~{\rm in}~\GO_i,~~i=1,2,\ldots, N \nonum
                 &=& \Be_0~{\rm elsewhere~outside}~\cup_{i}\GO_i,
\eeqa{3.4}
and periodically extend it. Then we get
\beqa  \lang\underline{\Be}_C\cdot\Gs_C\underline{\Be}_C\rang & = & p_n\Gs_2\Be_0\cdot\Be_0 \nonum
&~& +(1-p_n)\sum_{i=1}^N\lang\Be(\Bx/a_i+\Bb_i)\cdot\Gs(\Bx/a_i+\Bb_i)\Be(\Bx/a_i+\Bb_i)\rang_{\GO_i} \nonum
&=& p_n\Gs_2\Be_0\cdot\Be_0+(1-p_n)\lang\Be(\Bx)\cdot\Gs(\Bx)\Be(\Bx)\rang_{\GO} \nonum
&=& p_n\Gs_2\Be_0\cdot\Be_0+(1-p_n)\Be_0\cdot\BGs^D\Be_0,
\eeqa{3.5}
where $\lang\cdot\rang_{\GO_i}$ denotes an average over the region $\GO_i$.
Combined with the variational inequality \eq{3.2} this implies the bound
\beq \Be_0\cdot\BGs^*_n\Be_0\leq p_n\Gs_2\Be_0\cdot\Be_0+(1-p_n)\Be_0\cdot\BGs^D\Be_0
\leq \Be_0\cdot\BGs^D\Be_0,
\eeq{3.6}
where we have used the inequality $\BGs^D\geq\Gs_2\BI$ implied by \eq{2.13}. Thus we get
\beq \BGs^*_n\leq \BGs^D.
\eeq{3.7}
This composite has a volume fraction $f_1'=(1-p_n)f_1$ of phase 1.
Thus any bound ``from below'' on the effective conductivity $\BGs^*_n$, applicable to composites
having a volume fraction $f_1'$ of phase 1, immediately translates into
bound  ``from below'' on $\BGs^D$. Now consider what happens as we increase $N$, inserting
more and more regions $\GO_i$, while leaving undisturbed the regions $\GO_i$ already in place,
so that $p_n\to 0$ as $n\to\infty$. We are assured that this is possible. Rescaled copies
of any shaped region can be packed to fill all space: see, for example, Theorem A.1
in \citeAPY{Benveniste:2003:NER}. Define
\beq \BGs^*=\lim_{n\to\infty}\BGs^*_n.
\eeq{3.7a}
We are assured this limit exists since if we change the geometry in some
small volume then the effective conductivity (assuming $\Gs_1$ and $\Gs_2$ are strictly positive and finite)
is perturbed only by a small amount (\citeAY{Zhikov:1994:HDO}).
We will call $\BGs^*$ the effective conductivity tensor of an assemblage of rescaled copies
of $\GO$ packed to fill all space. Then \eq{3.7} implies
\beq \BGs^*\leq \BGs^D, \eeq{3.7b}
which is essentially the bound of \citeAPY{Huet:1990:AVC} applied to this assemblage.
Assume the bound is continuous with respect
to $f_1'$ at the point $f_1'=f_1$, as expected. Then taking the limit $n\to\infty$ the 
``lower bound'' on the effective tensor of composites having
volume fraction $f_1$ must also be a lower bound on $\BGs^D$.

In particular, the harmonic mean bound 
$\BGs^*\geq\lang\Gs^{-1}\rang^{-1}\BI$ translates into the elementary bound 
$\BGs^D\geq\lang\Gs^{-1}\rang^{-1}\BI$ of Nemat-Nasser and Hori, obtained before. 
Additionally, in our two-phase composite, the effective conductivity $\BGs^*$ satisfies
the Lurie-Cherkaev-Murat-Tartar bound (Lurie and Cherkaev \citeyearNP{Lurie:1982:AEC}, \citeyearNP{Lurie:1984:EEC}; \citeAY{Murat:1985:CVH}; \citeAY{Tartar:1985:EFC})
\beq f_1{\rm Tr}[(\BGs^*-\Gs_2\BI)^{-1}]\leq d/(\Gs_1-\Gs_2)+f_2/\Gs_2,
\eeq{3.8}
[which are a generalization of the bounds of \citeAPY{Hashin:1962:VAT}] where $d=2,3$ is the dimensionality
of the composite. Since $\BGs^D\geq\BGs^*\geq \Gs_2\BI$ it follows that 
\beq (\BGs^*-\Gs_2\BI)^{-1}\geq (\BGs^D-\Gs_2\BI)^{-1}, \eeq{3.9}
and so \eq{3.8} implies the new bound
\beq  f_1{\rm Tr}[(\BGs^D-\Gs_2\BI)^{-1}]\leq d/(\Gs_1-\Gs_2)+f_2/\Gs_2.
\eeq{3.10}
By multiplying this inequality by $\Gs_2^2$ and adding $df_1\Gs_2$ to both sides we see that it  
can be rewritten in the equivalent form 
\beq f_1{\rm Tr}[(\Gs_2^{-1}\BI-(\BGs^D)^{-1})^{-1}]\leq d/(\Gs_2^{-1}-\Gs_1^{-1})-(d-1)f_2\Gs_2.
\eeq{3.10a}
As $d^2/{\rm Tr}(\BA)\leq {\rm Tr}(\BA^{-1})$ for any positive definite matrix $\BA$ we also obtain
the weaker bound
\beq \frac{1}{d}{\rm Tr}[(\BGs^D)^{-1}]\leq \Gs_2^{-1}-\frac{f_1 d}{d/(\Gs_2^{-1}-\Gs_1^{-1})-(d-1)f_2\Gs_2},
\eeq{3.11}
which is a particular case of the universal bounds first derived by 
Nemat-Nasser and Hori (\citeyearNP{Nemat-Nasser:1993:MOP}, \citeyearNP{Nemat-Nasser:1995:UBO}), see equation (5.4.9) in their 1995 paper,
obtained under the assumption that $\GO$ is ellipsoidal or parallelpipedic.
(which we see is not needed).

If one is interested in bounds on the volume fraction $f_1$ then \eq{3.10} implies the upper bound
\beq f_1\leq \frac{1/\Gs_2+d/(\Gs_1-\Gs_2)}{1/\Gs_2+{\rm Tr}[(\BGs^D-\Gs_2\BI)^{-1}]}.
\eeq{3.12}

To obtain lower bounds on $f_1$, we consider the same periodic composite
and apply the dual variational inequality
\beq \Bj_0\cdot(\BGs^*_n)^{-1}\Bj_0\leq \lang\underline{\Bj}_C\cdot\Gs_C^{-1}\underline{\Bj}_C\rang
\eeq{3.13}
valid for any trial current field $\underline{\Bj}_C$ satisfying
\beq \Div\underline{\Bj}_C=0,\quad\underline{\Bj}_C~{\rm periodic},\quad
\lang\underline{\Bj}_C\rang=\Bj_0.
\eeq{3.14}
Letting $\Bj(\Bx)$ denote the current field within $\GO$ when the special
Neumann boundary conditions \eq{1.7} are applied, we may take in the unit cell
\beqa \underline{\Bj}_C(\Bx) &=& \Bj(\Bx/a_i+\Bb_i)~{\rm in}~\GO_i,~~i=1,2,\ldots, N \nonum
                 &=& \Bj_0~{\rm elsewhere~outside}~\cup_{i}\GO_i,
\eeqa{3.15}
and periodically extend it. 
Substituting this trial field in \eq{3.13} gives the bound 
\beq (\BGs^*_n)^{-1}\leq p_n\Gs_2^{-1}\BI+(\BGs^N)^{-1}, \eeq{3.15a}
which in the limit $n\to\infty$ implies
\beq  \BGs^*\geq \BGs^N, \eeq{3.16}
which is essentially the bound of \citeAPY{Huet:1990:AVC} applied to the
assemblage of rescaled copies of $\GO$ packed to fill all space.

Thus any bound ``from above'' on the effective conductivity $\BGs^*_n$ of composites having a volume 
fraction $f_1'$ immediately 
translates into a bound  ``from above'' on $(p_n\Gs_2^{-1}\BI+(\BGs^N)^{-1})^{-1}$. Taking
the limit $n\to\infty$
and assuming continuity of the bound at $f_1'=f_1$ the ``upper bound'' 
on the effective tensor of composites having volume fraction $f_1$ must also be an upper bound
on $\BGs^N$.
In particular, the other Murat-Tartar-Lurie-Cherkaev bound 
\beq f_2{\rm Tr}[(\Gs_1\BI-\BGs^*)^{-1}]\leq d/(\Gs_1-\Gs_2)-f_1/\Gs_1,
\eeq{3.17}
implies
\beq f_2{\rm Tr}[(\Gs_1\BI-\BGs^N)^{-1}]\leq d/(\Gs_1-\Gs_2)-f_1/\Gs_1. 
\eeq{3.18}
Again using the inequality $d^2/{\rm Tr}(\BA)\leq {\rm Tr}(\BA^{-1})$ for $\BA>0$, we obtain
the weaker bound
\beq  \frac{1}{d}{\rm Tr}(\BGs^N)\leq \Gs_1-\frac{f_2 d}{d/(\Gs_1-\Gs_2)-f_1/\Gs_1}
\eeq{3.20}
which is a particular case of the universal bounds derived by Nemat-Nasser and Hori (\citeyearNP{Nemat-Nasser:1993:MOP}, \citeyearNP{Nemat-Nasser:1995:UBO}), see 
equation (5.3.11) in their 1995 paper,
obtained under the assumption that $\GO$ is ellipsoidal or parallelpipedic
(which we see is not needed).

From \eq{3.18} we directly obtain the volume fraction bound
\beq f_2\leq \frac{d/(\Gs_1-\Gs_2)-1/\Gs_1}{\{{\rm Tr}[(\Gs_1\BI-\BGs^N)^{-1}]\}-1/\Gs_1},
\eeq{3.21}
giving a lower bound on the volume fraction $f_1=1-f_2$. 

In the asymptotic limit as the volume fraction goes to zero
the volume fraction bounds \eq{3.12} and \eq{3.21} reduce to those of Capdeboscq and Vogelius (\citeyearNP{Capdeboscq:2003:OAE}, \citeyearNP{Capdeboscq:2004:RSR}),
as shown in the two dimensional case by \citeAPY{Kang:2011:SBV}. The paper of Kang, Kim and Milton also tests the bounds numerically, and their (two-dimensional) results 
show the bound \eq{3.12} is typically close to the actual volume fraction for a variety of inclusions of phase 1 in a matrix of phase 2. Similarly we can expect
that the bound \eq{3.21} will be typically close to the actual volume fraction for a variety of inclusions of phase 2 in a matrix of phase 1.

\section{Coupled bounds in two-dimensions}
\setcounter{equation}{0}
The tensors $\BGs^D$ and $\BGs^N$ obviously depend on $\Gs_1$ and $\Gs_2$, i.e. 
$\BGs^D=\BGs^D(\Gs_1,\Gs_2)$ and $\BGs^N=\BGs^N(\Gs_1,\Gs_2)$. Let us assume we have
measurements of these tensors for an additional pair of conductivities $(k_1, k_2)$,
(which could be obtained, say from thermal, magnetic permeability, or diffusivity
measurements) and let $\Bk^D$ and $\Bk^N$ denote these tensors,
\beq \Bk^D=\BGs^D(k_1,k_2),\quad \Bk^N=\BGs^N(k_1,k_2).
\eeq{4.1}
We still let $\BGs^D$ and $\BGs^N$ denote the tensors associated with the first
pair of conductivities $(\Gs_1, \Gs_2)$, with $\Gs_1>\Gs_2$. From \eq{3.7b} and
\eq{3.16} we have the inequalities
\beqa  \Gs_2\BI\leq\BGs^N\leq\BGs^*\leq \BGs^D\leq \Gs_1\BI, \nonum
k^-\leq \Bk^N\leq\Bk^*\leq \Bk^D\leq k^+\BI,
\eeqa{4.2}
where $k^-=\min\{k_1,k_2\}$ and $k^+=\max\{k_1,k_2\}$ and
$\Bk^*$ is the effective conductivity the composite considered
in the previous section when  $\Gs_1$ and $\Gs_2$ are replaced by $k_1$ and $k_2$. (It can
easily be checked that these inequalities still hold if $k_2>k_1$.) 

For two dimensional conductivity from duality 
(\citeAY{Keller:1964:TCC}; \citeAY{Dykhne:1970:CTD})
we know the functions 
$\BGs^D=\BGs^D(\Gs_1,\Gs_2)$ and $\BGs^N=\BGs^N(\Gs_1,\Gs_2)$ satisfy
\beqa \BGs^D(\Gs_2,\Gs_1)& = &\Gs_1\Gs_2\BR_\perp^T[\BGs^N(\Gs_1,\Gs_2)]^{-1}\BR_\perp, \nonum
 \BGs^N(\Gs_2,\Gs_1)& = &\Gs_1\Gs_2\BR_\perp^T[\BGs^D(\Gs_1,\Gs_2)]^{-1}\BR_\perp,
\eeqa{4.3}
where 
\beq \BR_\perp=\pmatrix{0 & 1 \cr -1 & 0}
\eeq{4.4}
is the matrix for a $90^\circ$ rotation.
So if we know these tensors for the conductivity pair $(k_1, k_2)$, we also know them for the 
conductivity pair $(k_2, k_1)$. Hence, by making such an interchange if necessary, we may assume without
loss of generality that $k_1>k_2$, i.e. that $k^+=k_1$ and $k^-=k_2$. Finally, by 
interchanging $k$ with $\Gs$ if necessary,
we may assume without loss of generality that 
\beq \Gs_1/\Gs_2\geq k_1/k_2>1. \eeq{4.5}

\begin{figure}[htbp]
\begin{center}
\epsfig{figure=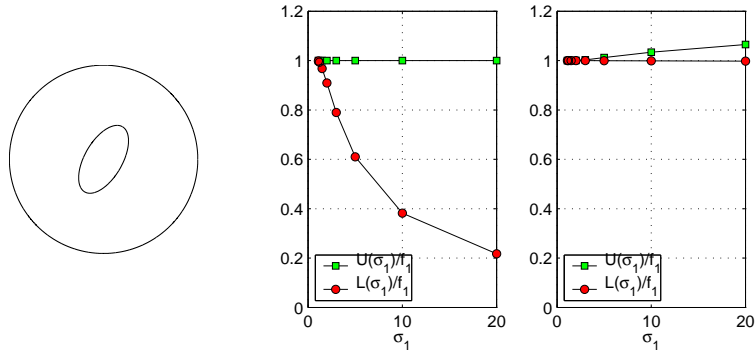,width=10cm}
\end{center}
\caption{The first figure shows the circular body $\GO$ containing an ellipse of phase 1 surrounded by phase 2.
The second figure shows the results for the bounds
\eq{3.12} and \eq{3.21} while the third figure shows the results 
for the bounds \eq{4.16}.
The bounds are for increasing $\sigma_1$, with $\sigma_2=1$
and (for the third figure) the pairs $(\sigma_1,k_1)$ are taken as
$(1.1,1.05)$, $(1.2,1.1)$, $(1.5,1.2)$, $(2,1.5)$, $(3,2)$, $(5,3)$, $(10,5)$ and $(20,10)$,
with $\sigma_2=k_2=1$.
Here $U(\Gs_1)$ and $L(\Gs_1)$ are the upper and 
lower bounds on the volume fraction, and the true volume fraction is $f_1=0.08$.
Figure supplied courtesy of Eunjoo Kim.}\label{3}
\end{figure}

\begin{figure}[htbp]
\begin{center}
\epsfig{figure=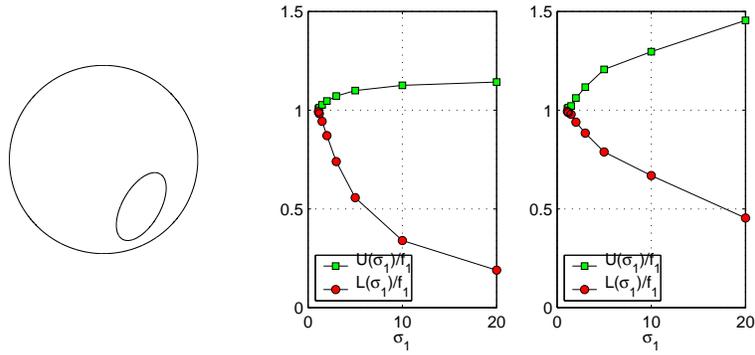,width=10cm}
\end{center}
\caption{The same as for figure \ref{3} but with the elliptical 
inclusion moved closer to the boundary of $\GO$. Figure supplied courtesy of Eunjoo Kim.}\label{4}
\end{figure}

\begin{figure}[htbp]
\begin{center}
\epsfig{figure=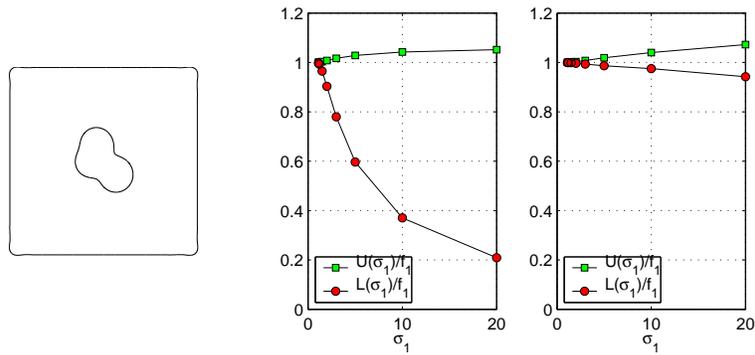,width=10cm}
\end{center}
\caption{The same as for figure \ref{3} but with a non-elliptical inclusion of phase 1 in a square region $\GO$. The true volume
fraction is $f_1=0.0673$. Figure supplied courtesy of Eunjoo Kim. }\label{5}
\end{figure}

Optimal bounds on all possible matrix pairs $(\BGs^*,\Bk^*)$  for composites having a prescribed
volume fraction $f_1$ of phase 1 have been derived by \citeAPY{Cherkaev:1992:ECB}, 
and extended to an arbitrary number of effective conductivity function values by  \citeAPY{Clark:1995:OBC}.
However it seems
difficult to extract bounds on $f_1$ from these optimal bounds. Instead we consider a
polycrystal checkerboard with conductivities 
\beq \BGs(\Bx)=\BR^T(\Bx)\BGs^*\BR(\Bx),\quad \Bk(\Bx)=\BR^T(\Bx)\Bk^*\BR(\Bx),
\quad {\rm with}~\BR^T(\Bx)\BR(\Bx)=\BI,
\eeq{4.6}
in which the rotation field $\BR(\Bx)$ is $\BI$ in the ``white squares'' and
$\BR_\perp$ in the ``black squares''. By a result of \citeAPY{Dykhne:1970:CTD} this material
has effective conductivities $(\Gs_*\BI,k_*\BI)$ where
\beq \Gs_*=\sqrt{\det\BGs^*},\quad k_*=\sqrt{\det\Bk^*}.
\eeq{4.7}
Now we replace the ``white squares'' by the limiting composite considered
in the previous section (with structure much smaller than the size of the
squares) and we replace the ``black squares'' by the limiting composite considered
in the previous section, rotated by $90^\circ$. The resulting material is an isotropic
composite of phases 1 and 2 and so the pair $(\Gs_*,k_*)$ satisfies the bounds
of \citeAPY{Milton:1981:BTO},
\beq u(k_*)\leq \Gs_* \leq v(k_*), \eeq{4.10}
which are attained when the composite is an assemblage of doubly coated disks, where
\beqa 
 v(k_*) & = &\Gs_1-\frac{2f_2\Gs_1(\Gs_1^2-\Gs_2^2)}{(f_2\Gs_1+f_1\Gs_2+\Gs_1)(\Gs_1+\Gs_2)
+(\Gs_1-\Gs_2)^2\Ga_1(k_*)}, \nonum
u(k_*) & \equiv &\Gs_2+
\frac{2f_1\Gs_2(\Gs_1^2-\Gs_2^2)}{(f_2\Gs_1+f_1\Gs_2+\Gs_2)(\Gs_1+\Gs_2)
+(\Gs_1-\Gs_2)^2\Ga_2(k_*)},
\eeqa{4.11}
and
\beqa
\Ga_1(k_*)& = &\frac{(k_1+k_2)[2f_2k_1(k_1-k_2)/(k_1-k_*)-(f_2k_1+f_1k_2+k_1)]}{(k_1-k_2)^2}, \nonum
\Ga_2(k_*)& = &\frac{(k_1+k_2)[2f_1k_2(k_1-k_2)/(k_*-k_2)-(f_2k_1+f_1k_2+k_2)]}{(k_1-k_2)^2}.
\eeqa{4.12}
Now for any two symmetric matrices $\BA$ and $\BB$ with $\BA\geq\BB>0$ we 
have $\BB^{-1/2}\BA\BB^{-1/2}\geq\BI$, and so $\det(\BB^{-1/2}\BA\BB^{-1/2})\geq 1$
implying $\det(\BA)>\det(\BB)$. Thus \eq{4.2} and \eq{4.7} imply
\beq \Gs_2\leq\Gs_N\leq\Gs_*\leq \Gs_D\leq\Gs_1, 
\quad k_2\leq k_N\leq k_*\leq k_D\leq k_1,
\eeq{4.13}
where we define
\beq \Gs_N=\sqrt{\det\BGs_N},\quad \Gs_D=\sqrt{\det\BGs_D},\quad
 k_N=\sqrt{\det\Bk_N},\quad k_D=\sqrt{\det\Bk_D}.
\eeq{4.13a}
The Hashin-Shtrikman bounds (\citeAY{Hashin:1962:VAT}; \citeAY{Hashin:1970:TCM}),
\beq k_1-\frac{2f_2k_1(k_1-k_2)}{f_2k_1+f_1k_2+k_1}\geq k_*
\geq  k_2+\frac{2f_1k_2(k_1-k_2)}{f_2k_1+f_1k_2+k_2},
\eeq{4.14}
imply that both $\Ga_1(k_*)$ and $\Ga_2(k_*)$ are non-negative. Hence the denominators
in \eq{4.11} are positive and so \eq{4.10} implies
\beqa 
(\Gs_1-\Gs_*)[(f_2\Gs_1+f_1\Gs_2+\Gs_1)(\Gs_1+\Gs_2)
+(\Gs_1-\Gs_2)^2\Ga_1(k_*)]\geq 2f_2\Gs_1(\Gs_1^2-\Gs_2^2),
\nonum
(\Gs_*-\Gs_2)[(f_2\Gs_1+f_1\Gs_2+\Gs_2)(\Gs_1+\Gs_2)
+(\Gs_1-\Gs_2)^2\Ga_2(k_*)]\geq 2f_1\Gs_2(\Gs_1^2-\Gs_2^2).
\eeqa{4.15}
Since $\Ga_1(k_D)\geq\Ga_1(k_*)$ and $\Ga_2(k_N)\geq\Ga_2(k_*)$, we get using
\eq{4.13},
\beqa
(\Gs_1-\Gs_N)[(f_2\Gs_1+f_1\Gs_2+\Gs_1)(\Gs_1+\Gs_2)
+(\Gs_1-\Gs_2)^2\Ga_1(k_D)]\geq 2f_2\Gs_1(\Gs_1^2-\Gs_2^2),
\nonum
(\Gs_D-\Gs_2)[(f_2\Gs_1+f_1\Gs_2+\Gs_2)(\Gs_1+\Gs_2)
+(\Gs_1-\Gs_2)^2\Ga_2(k_N)]\geq 2f_1\Gs_2(\Gs_1^2-\Gs_2^2).
\nonum ~
\eeqa{4.16}
As $\Ga_1(k_D)$ and $\Ga_2(k_N)$ depend linearly on $f_1$ and $f_2=1-f_1$,
the equations \eq{4.16} readily yield bounds on the volume fraction. Eunjoo Kim
has used an integral equation solver [as described by \citeAPY{Kang:2011:SBV}]
to compare the bounds \eq{4.16} with the bounds \eq{3.12} and \eq{3.21}. Her results
are presented in figures \ref{3}, \ref{4}, and \ref{5}. More numerical results testing
the bounds \eq{3.12} and \eq{3.21} are in the paper by \citeAPY{Kang:2011:SBV}.

\section{Coupled bounds in three-dimensions}
\setcounter{equation}{0}

We can also derive coupled bounds in three dimensions. Let us assume the phases
have been labeled so that
\beq \Gs_1 k_1\geq \Gs_2 k_2, \quad {\rm i.e.}~\Gs_1/\Gs_2\geq k_2/k_1,
\eeq{4.17}
and by interchanging $\Gs$ with $k$ if necessary let us assume
\beq \Gs_1/\Gs_2\geq k_1/k_2.
\eeq{4.18}
These two inequalities imply $\Gs_1/\Gs_2>1$ as before. We want to use the
inequalities \eq{4.2} to derive bounds on the volume fraction. As in the 
two-dimensional case the idea is to first construct an isotropic
polycrystal, where the polycrystal has the conductivities \eq{4.6}
in which the rotation field $\BR(\Bx)$ is constant within grains
which we take to be spheres. These spheres fill all space, and the crystal
orientation varies randomly from sphere to sphere so that the composite
has isotropic conductivities $(\Gs_*\BI,k_*\BI)$. We use the effective medium formula
(\citeAY{Stroud:1975:GEM}; \citeAY{Helsing:1991:ECA})
which gives 
\beq \Gs_*=g(\BGs_*), \quad k_*=g(\Bk_*),
\eeq{4.19}
where for any positive definite symmetric $3\times 3$ matrix $\BA$, $g=g(\BA)$ is taken to be the
unique positive root of
\beq \frac{\Gl_1-g}{\Gl_1+2g}+\frac{\Gl_2-g}{\Gl_2+2g}+\frac{\Gl_3-g}{\Gl_3+2g}=0,
\eeq{4.20}
in which $\Gl_1$, $\Gl_2$, and $\Gl_3$ are the eigenvalues of $\BA$. This effective
medium formula is realizable (\citeAY{Milton:1985:TCP}; \citeAY{Avellaneda:1987:IHD})
in the sense that it corresponds to a limiting composite
of spherical grains with hierarchical structure (where any pair of grains of 
comparable size are well separated from each other, relative to their diameter). 
Note that the left hand side of side of \eq{4.20} increases if any of the eigenvalues
$\Gl_i$ increase, and decreases if $g$ increases. So $g(\BA)$ must increase if
any or all of the eigenvalues of $\BA$ increase. It follows that $g(\BB)\geq g(\BA)$
if $\BB\geq\BA>0$. Hence the inequalities \eq{4.2} imply
\beq \Gs_2\leq\Gs_N\leq\Gs_*\leq \Gs_D\leq\Gs_1, 
\quad k^-\leq k_N\leq k_*\leq k_D\leq k^+,
\eeq{4.20a}
where now
\beq  \Gs_N=g(\BGs_N),\quad \Gs_D=g(\BGs_D),\quad k_N=g(\Bk_N),\quad k_D=g(\Bk_D).
\eeq{4.20b}

We next replace the material in each sphere by the appropriately oriented 
limiting composite considered in the previous section (with structure much 
smaller than the sphere diameter) to obtain a two-phase isotropic composite with
$(\Gs_*,k_*)$ as its conductivities. Thus $\Gs_*$ must satisfy the upper bound of Bergman (\citeyearNP{Bergman:1976:VBS},\citeyearNP{Bergman:1978:DCC})
\beq
\Gs_*\leq  f_1\Gs_1+f_2\Gs_2
-\frac{f_1f_2(\Gs_1-\Gs_2)^2}{3\Gs_2+(\Gs_1-\Gs_2)\Gg(k_*)},
\eeq{4.23}
where
\beq \Gg(k_*)=\frac{f_1f_2(k_1-k_2)}{f_1k_1+ f_2k_2 -k_*}-\frac{3k_2}{k_1-k_2},
\eeq{4.24}
and the lower bound
\beq \Gs_*\geq \Gs_2+
\frac{3f_1\Gs_2(\Gs_1-\Gs_2)(\Gs_2+2\Gs_1)}{(f_2\Gs_1+f_1\Gs_2+2\Gs_2)(\Gs_2+2\Gs_1)
+(\Gs_1-\Gs_2)^2\Gb(k_*)},
\eeq{4.21}
where
\beq
\Gb(k_*) = \frac{(k_2+2k_1)[3f_1k_2(k_1-k_2)/(k_*-k_2)-(f_2k_1+f_1k_2+2k_2)]}{(k_1-k_2)^2}.
\eeq{4.22}
This lower bound was first conjectured by \citeAPY{Milton:1981:BTO}. A proof was proposed
by \citeAPY{Avellaneda:1988:ECP} which was corrected by \citeAPY{Nesi:1991:MII} and \citeAPY{Zhikov:1991:EHM}.

The lower bound \eq{4.21} is sharp, being attained for two-phase assemblages of doubly
coated spheres (\citeAY{Milton:1981:BTO}). The upper bound \eq{4.23} is attained at $5$ values of $\Gg(k_*)$
namely when $\Gg(k_*)=f_2, 3f_2/2, 3f_2, 3-3f_1/2,$ and $3-f_1$ (\citeAY{Milton:1981:BCP}).

The Hashin-Shtrikman bound (\citeAY{Hashin:1962:VAT}),
\beq (k_*-k_2)/(k_1-k_2)\geq 3f_1k_2/(f_2k_1+f_1k_2+2k_2),
\eeq{4.25}
implies that $\Gb(k_*)$ is non-negative.  Hence the denominator
in \eq{4.21} is positive and so the inequality implies
\beq (\Gs_*-\Gs_2)[(f_2\Gs_1+f_1\Gs_2+2\Gs_2)(\Gs_2+2\Gs_1)
+(\Gs_1-\Gs_2)^2\Gb(k_*)]\geq 3f_1\Gs_2(\Gs_1-\Gs_2)(\Gs_2+2\Gs_1).
\eeq{4.26}
The Hashin-Shtrikman bounds can also be rewritten in the form
\beq
f_2k_1+f_1k_2+2k^- \leq \frac{f_1f_2(k_1-k_2)^2}{f_1k_1+ f_2k_2 -k_*}\leq f_2k_1+f_1k_2+2k^+.
\eeq{4.27}
These inequalities imply
$\Gg(k_*)$ lies between $f_2$ and $3-f_1$. Hence the denominator in \eq{4.23} is positive and
the inequality can be rewritten as 
\beq
(f_1\Gs_1+f_2\Gs_2-\Gs_*)(3\Gs_2+(\Gs_1-\Gs_2)\Gg(k_*))\geq f_1f_2(\Gs_1-\Gs_2)^2.
\eeq{4.28}
When $k_1\geq k_2$ \eq{4.20a} implies  $\Gb(k_N)\geq\Gb(k_*)$ and $\Gg(k_D)\geq\Gg(k_*)$, and 
hence 
\beqa
(\Gs_D-\Gs_2)[(f_2\Gs_1+f_1\Gs_2+2\Gs_2)(\Gs_2+2\Gs_1)
+(\Gs_1-\Gs_2)^2\Gb(k_N)]& \geq & 3f_1\Gs_2(\Gs_1-\Gs_2)(\Gs_2+2\Gs_1),\nonum
(f_1\Gs_1+f_2\Gs_2-\Gs_N)(3\Gs_2+(\Gs_1-\Gs_2)\Gg(k_D))& \geq & f_1f_2(\Gs_1-\Gs_2)^2. \nonum
\eeqa{4.29}
On the other hand when $k_1\leq k_2$ then \eq{4.20a} implies $\Gb(k_D)\geq\Gb(k_*)$ and
$\Gg(k_N)\geq\Gg(k_*)$, and hence 
\beqa
(\Gs_D-\Gs_2)[(f_2\Gs_1+f_1\Gs_2+2\Gs_2)(\Gs_2+2\Gs_1)
+(\Gs_1-\Gs_2)^2\Gb(k_D)]& \geq & 3f_1\Gs_2(\Gs_1-\Gs_2)(\Gs_2+2\Gs_1),\nonum
(f_1\Gs_1+f_2\Gs_2-\Gs_N)(3\Gs_2+(\Gs_1-\Gs_2)\Gg(k_N))&\geq& f_1f_2(\Gs_1-\Gs_2)^2. \nonum
\eeqa{4.30}
Since $\Gb(k_N)$ and $\Gb(k_D)$ depend linearly on the volume fractions $f_1$ and $f_2=1-f_1$, 
the first inequalities
in \eq{4.29} and \eq{4.30} also depend linearly on the volume fraction and easily yield
bounds on the volume fraction. On the other hand, finding bounds on the volume fraction
from the second inequalities in \eq{4.29} and \eq{4.30}, involves solving a cubic equation in $f_1$. So instead
of analytically computing the roots of this cubic it is probably better to numerically
search for the range of values of $f_1$ where the second inequalities in \eq{4.29} and \eq{4.30}
are satisfied.

\section{Bounds for elasticity}
\setcounter{equation}{0}

Let us consider solutions to the linear elasticity equations 
\beq \BGj(\Bx)=\BCC(\Bx)\BGe(\Bx),\quad\Div\BGt=0,\quad\BGe=(\Grad\Bu+(\Grad\Bu)^T)/2,
\eeq{5.1}
within $\GO$, where $\Bu(\Bx)$, $\BGe(\Bx)$ and $\BGj(\Bx)$, are the displacement
field, strain field, and stress field, and $\BCC(\Bx)$ is the fourth
order elasticity tensor field
\beq \BCC(\Bx)=\Gc(\Bx)\BCC^1+(1-\Gc(\Bx))\BCC^2,
\eeq{5.2}
in which $\BCC^1$ and $\BCC^2$ are the elasticity tensors of the phases, assumed
to be isotropic with elements,
\beq \CC_{ijk\ell}^h
=\mu_h(\Gd_{ik}\Gd_{j\ell}+\Gd_{i\ell}\Gd_{jk})+(\Gk_h-2\mu_h/d)\Gd_{ij}\Gd_{k\ell},\quad h=1,2,
\eeq{5.3}
in which $d=2$ or 3 is the dimensionality, and 
$\Gm_1,\Gm_2$ and $\Gk_1,\Gk_2$ are the shear and bulk moduli of the
two phases. From boundary information on the displacement $\Bu_0(\Bx)=\Bu(\Bx)$
and traction $\Bf(\Bx)=\BGj(\Bx)\cdot\Bn$
we can immediately determine, using integration by parts, 
volume averages such as
\beqa \lang \BGe:\BGj\rang &= & \frac{1}{|\GO|}\int_{\partial\GO}\Bu\cdot\Bf, \nonum
      \lang \BGe\rang & = & \frac{1}{|\GO|}\int_{\partial\GO}(\Bn\Bu^{T}+\Bu\Bn^{T})/2, \nonum
      \lang \BGj\rang & = & \frac{1}{|\GO|}\int_{\partial\GO}\Bx\Bf^T,
\eeqa{5.3a}
in which $":"$ denotes a contraction of two indices.

There are two natural sets of boundary conditions. 
For any symmetric matrix $\BGe_0$ we could
prescribe the special Dirichlet boundary conditions 
\beq \Bu(\Bx)=\BGe_0\Bx,\quad {\rm for}~\Bx\in\Md\GO, 
\eeq{5.4}
and measure $\BGj_0=\lang\BGj\rang$. Here, according to \eq{5.3a}, $\BGe_0$ equals $\lang\BGe\rang$. Since
$\BGj_0$ is linearly related to $\BGe_0$ we can write
\beq \BGj_0=\BCC^D\BGe_0,
\eeq{5.5}  
which defines the elasticity tensor $\BGs^D$ ($D$ for Dirichlet). Alternatively
for any symmetric matrix $\BGj_0$ we could
prescribe the special Neumann boundary conditions 
\beq \BGj(\Bx)\cdot\Bn=\BGj_0\cdot\Bn,\quad {\rm for}~\Bx\in\Md\GO, 
\eeq{5.6}
and measure $\BGe_0=\lang\BGe\rang$. Here, according to \eq{5.3a}, $\BGj_0$ equals $\lang\BGj\rang$. Since
$\BGe_0$ is linearly related to $\BGj_0$ we can write
\beq \BGe_0=(\BCC^N)^{-1}\BGj_0, 
\eeq{5.7}  
which defines the elasticity tensor $\BGs^N$ ($D$ for Dirichlet). It is easy to check
that $\BCC^D$ and $\BCC^N$ satisfy all the usual symmetries of elasticity tensors.

Directly analogous to \eq{2.13} we have the bounds 
\beq  \lang\BCC^{-1}\rang^{-1}\leq\BCC^D\leq \lang\BCC\rang,\quad\quad
 \lang\BCC^{-1}\rang^{-1}\leq\BCC^N\leq \lang\BCC\rang
\eeq{5.8}
of \cite{Nemat-Nasser:1993:MOP}, and directly analogous to \eq{2.14} 
for any boundary condition (not just the
special boundary conditions \eq{5.4} and \eq{5.6}) we have the bounds
\beq \lang\BGe\cdot\BGj\rang \geq \BGe_0\cdot\BGs^N\BGe_0,\quad
\lang\BGe\cdot\BGj\rang \geq \BGj_0(\BGs^D)^{-1}\BGj_0,
\eeq{5.9}
where $\BGe_0=\lang\BGe\rang$ and $\BGj_0=\lang\BGj\rang$,
due to Willis in 1989 private communication to Nemat-Nasser and Hori and presented by \citeAPY{Nemat-Nasser:1993:MOP}.

Also directly analogous to \eq{3.7b} and \eq{3.16} we have the bounds
\beq \BCC^N\leq\BCC^*\leq \BCC^D,
\eeq{5.10}
where $\BCC^*$ is the effective elasticity tensor of any assemblage of
rescaled copies of $\GO$ packed to fill all space. These are 
essentially the bounds of \citeAPY{Huet:1990:AVC} applied to this assemblage. Thus ``lower
bounds'' on $\BCC^*$ directly give ``lower bounds'' on $\BCC^D$ and
``upper bounds'' on $\BCC^*$ directly give ``upper bounds'' on $\BCC^N$.
In particular, in two dimensions lower and upper bounds on $\BGe_0\cdot\BCC^*\BGe_0$ 
have been obtained by \citeAPY{Gibiansky:1984:DCPa} (for the equivalent plate equation) and also by \citeAPY{Allaire:1993:EOB}.
Assuming that the phases have been labeled so that $\Gm_1\geq\Gm_2$ ($\Gk_1-\Gk_2$ could be either positive
or negative)
and letting $\Ge_1$ and $\Ge_2$ denote the two eigenvalues $\BGe_0$,
the bounds imply
\beqa
&~& \BGe_0\cdot\BCC^D\BGe_0  \geq
(\Ge_1+\Ge_2)^2/(f_1/\Gk_1+f_2/\Gk_2)
+(\Ge_1-\Ge_2)^2/(f_1/\Gm_1+f_2/\Gm_2), \nonum
&~& \quad \quad {\rm if~~} |\Gk_1-\Gk_2|(f_1\Gm_2+f_2\Gm_1)|\Ge_1+\Ge_2|\leq
 |\Gm_1-\Gm_2|(f_1\Gk_2+f_2\Gk_1)|\Ge_1-\Ge_2|; \nonum
&~& \nonum &~&
\BGe_0\cdot\BCC^D\BGe_0 \geq (\Ge_1+\Ge_2)^2(f_1\Gk_1+f_2\Gk_2)
+(\Ge_1-\Ge_2)^2(f_1\Gm_1+f_2\Gm_2) \nonum
&~&~~~~~~~~~~~~~~~~~~~-f_1f_2\frac{[|\Gk_1-\Gk_2||\Ge_1+\Ge_2|+|\Gm_1-\Gm_2||\Ge_1-\Ge_2|]^2}
{f_1(\Gm_2+\Gk_2)+f_2(\Gm_1+\Gk_1)}, \nonum
&~&\quad \quad {\rm if~~} (\Gm_2+f_1\Gk_2+f_2\Gk_1)|\Ge_1-\Ge_2| \geq
f_2|\Gk_1-\Gk_2||\Ge_1+\Ge_2| \nonum
&~& \quad \quad{\rm and~~}  |\Gk_1-\Gk_2|(f_1\Gm_2+f_2\Gm_1)|\Ge_1+\Ge_2|\geq
 |\Gm_1-\Gm_2|(f_1\Gk_2+f_2\Gk_1)|\Ge_1-\Ge_2|; \nonum
&~& \nonum &~&
 \BGe_0\cdot\BCC^D\BGe_0 \geq
\Gm_2(\Ge_1-\Ge_2)^2+
\frac{\Gk_1\Gk_2+\Gm_2(f_1\Gk_1+f_2\Gk_2)}{\Gm_2+f_1\Gk_2+f_2\Gk_1}(\Ge_1+\Ge_2)^2, \nonum
&~ & \quad \quad{\rm if~~} (\Gm_2+f_1\Gk_2+f_2\Gk_1)|\Ge_1-\Ge_2| \leq
f_2|\Gk_1-\Gk_2||\Ge_1+\Ge_2|;
\eeqa{5.12}
and
\beqa
&~& \BGe_0\cdot\BCC^N\BGe_0  \leq (\Ge_1+\Ge_2)^2(f_1\Gk_1+f_2\Gk_2)
+(\Ge_1-\Ge_2)^2(f_1\Gm_1+f_2\Gm_2) \nonum
&~&~~~~~~~~~~~~~~~~~~~-f_1f_2\frac{[|\Gk_1-\Gk_2||\Ge_1+\Ge_2|-|\Gm_1-\Gm_2||\Ge_1-\Ge_2|]^2}
{f_1(\Gm_2+\Gk_2)+f_2(\Gm_1+\Gk_1)}, \nonum
&~&\quad \quad {\rm if~~~} f_1|\Gk_1-\Gk_2||\Ge_1+\Ge_2|\leq (\Gm_1+f_1\Gk_2+f_2\Gk_1)|\Ge_1-\Ge_2| \nonum
&~&\quad \quad {\rm and~~} f_+|\Gm_1-\Gm_2||\Ge_1-\Ge_2|\leq (\Gk_++f_1\Gm_2+f_2\Gm_1)|\Ge_1+\Ge_2|; \nonum
&~& \nonum &~&
\BGe_0\cdot\BCC^N\BGe_0  \leq \Gm_1(\Ge_1-\Ge_2)^2+\frac{\Gk_1\Gk_2+\Gm_1(f_1\Gk_1+f_2\Gk_2)}{\Gm_1+f_1\Gk_2+f_2\Gk_1}(\Ge_1+\Ge_2)^2, \nonum
&~&\quad \quad {\rm if~~~} f_1|\Gk_1-\Gk_2||\Ge_1+\Ge_2|\geq (\Gm_1+f_1\Gk_2+f_2\Gk_1)|\Ge_1-\Ge_2|; \nonum
&~& \nonum &~&
\BGe_0\cdot\BCC^N\BGe_0  \leq \Gk_+(\Ge_1+\Ge_2)^2+\frac{\Gm_1\Gm_2+\Gk_+(f_1\Gm_1+f_2\Gm_2)}{\Gk_++f_1\Gm_2+f_2\Gm_1}(\Ge_1-\Ge_2)^2, \nonum
&~&\quad \quad {\rm if~~~} f_+|\Gm_1-\Gm_2||\Ge_1-\Ge_2|\geq (\Gk_++f_1\Gm_2+f_2\Gm_1)|\Ge_1+\Ge_2|,
\eeqa{5.12a}
where $\Gk_+$ is the maximum of $\Gk_1$ and $\Gk_2$ and $f_+$ is the volume fraction of the material corresponding to $\Gk_+$.

The corresponding three-dimensional bounds follow directly from
\eq{5.10} and the bounds of \citeAPY{Allaire:1993:OBE}, but are not so explicit. Assuming
that the Lame moduli
\beq \Gl_1=\Gk_1-2\Gm_1/3~~{\rm and}~~\Gl_2=\Gk_2-2\Gm_2/3 
\eeq{5.13}
of both phases are positive, and that the bulk and shear moduli of the two phases are well-ordered with
\beq \Gk_1>\Gk_2>0 {\rm ~~and~~}\Gm_1>\Gm_2>0, 
\eeq{5.13a}
these bounds are
\beqa
\BGe_0:\BCC^D\BGe_0 & \geq & \BGe_0:\BCC_2\BGe_0
+f_1\max_{\BGn}[2\BGe_0:\BGn-\BGn:(\BCC_1-\BCC_2)^{-1}\BGn-f_2g(\BGn)], \nonum
\BGe_0:\BCC^N\BGe_0 & \geq & \BGe_0:\BCC_1\BGe_0
+f_2\min_{\BGn}[2\BGe_0:\BGn+\BGn:(\BCC_1-\BCC_2)^{-1}\BGn-f_1h(\BGn)],
\eeqa{5.14}
where $g(\BGn)$ and $h(\BGn)$ are function of the eigenvalues $\Gn_1, \Gn_2$, and $\Gn_3$ 
of the symmetric matrix $\BGn$.
Assuming that
these are labeled with
\beq \Gn_1\leq\Gn_2\leq\Gn_3, 
\eeq{5.15}
we have
\beqa
g(\BGn)&=&
\frac{(\Gn_1-\Gn_3)^2}{4\Gm_2}+\frac{(\Gn_1+\Gn_3)^2}{4(\Gl_2+\Gm_2)}  ~~{\rm if}~~
\Gn_3\geq\frac{\Gl_2+2\Gm_2}{2(\Gl_2+\Gm_2)}(\Gn_1+\Gn_3)\geq\Gn_1, \nonum
g(\BGn)&=&
\frac{\Gn_1^2}{\Gl_2+2\Gm_2}  ~~{\rm if}~~
\Gn_1>\frac{\Gl_2+2\Gm_2}{2(\Gl_2+\Gm_2)}(\Gn_1+\Gn_3), \nonum
g(\BGn)&=&
\frac{\Gn_3^2}{\Gl_2+2\Gm_2}  ~~{\rm if}~~
\Gn_3<\frac{\Gl_2+2\Gm_2}{2(\Gl_2+\Gm_2)}(\Gn_1+\Gn_3),
\eeqa{5.16}
and
\beq h(\BGn)=\frac{1}{\Gl_1+2\Gm_1}\min\{\Gn_1^2,\Gn_2^2,\Gn_3^2\}.
\eeq{5.17}

The bounds \eq{5.12}, \eq{5.12a} and \eq{5.14} can be used in an inverse way to bound
the volume fraction $f_1=1-f_2$, for a single experiment when for special Dirichlet conditions
$\BGe_0$ is prescribed and $\BGj_0$ ($=\BCC^D\BGe_0$) is measured, or when for special Neumann conditions
$\BGj_0$ is prescribed and $\BGe_0$ ($=\BCC^N\BGe_0$) is measured. \citeAPY{Allaire:1993:OBE} also derive
bounds on the complementary energy and these imply
\beq \BGj_0:(\BCC^N)^{-1}\BGj_0  \geq  \BGj_0:\BCC_1^{-1}\BGj_0
+f_2\max_{\BGz}[2\BGj_0:\BGz-\BGz:(\BCC_2^{-1}-\BCC_1^{-1})^{-1}\BGz-f_1\BGz:\BCC_1\BGz+f_1h(\BCC_1\BGz)], \nonum
\eeq{5.17aa}
and
\beq
\BGj_0:(\BCC^D)^{-1}\BGj_0  \leq  \BGj_0:\BCC_2^{-1}\BGj_0
+f_1\min_{\BGz}[2\BGj_0:\BGz+\BGz:(\BCC_2^{-1}-\BCC_1^{-1})^{-1}\BGz-f_2\BGz:\BCC_2\BGz+f_2g(\BCC_2\BGz)].
\eeq{5.17ab}

The bound in \eq{5.17aa} is particularly useful when $\BGj_0=-p\BI$, corresponding to immersing the body $\GO$ in a fluid
with pressure $p$. Then from measurements of the resulting volume change of the body one can determine $\BGj_0:(\BCC^N)^{-1}\BGj_0=-p\Tr\BGe_0$.
Let us assume $\Gl_1>0$ and set $\BGz=\Ga\BI+\BA$, with $\BA$ being a trace free matrix with eigenvalues $a_1$, $a_2$ and $a_3$. Then 
we have $\BGn=\BCC_1\BGz=2\Gm_1(k\BI+\BA)$ where $k=\Ga[1+3\Gl_1/(2\Gm_1)]$. Substitution gives
\beqa &~&[\BGz:\BCC_1\BGz-h(\BCC_1\BGz)]-\Ga^2[\BI:\BCC_1\BI-h(\BCC_1\BI)] \nonum
&~&~=2\Gm_1\left[a_1^2+a_2^2+a_3^2-\frac{2\Gm_1}{\Gl_1+2\Gm_1}\min\{(k+a_1)^2-k^2,(k+a_2)^2-k^2,(k+a_3)^2-k^2\}\right] \nonum
&~&~\geq 2\Gm_1[a_1^2+a_2^2+a_3^2-\min\{2a_1k+a_1^2,2a_2k+a_2^2,2a_3k+a_3^2\}],
\eeqa{5.17ac}
which is surely positive since $\min\{2a_1k+a_1^2,2a_2k+a_2^2,2a_3k+a_3^2\}\leq a_j^2$ where $j$ is such that $ka_j$ is non-positive. (Note that
$a_1$, $a_2$ and $a_3$ cannot all have the same sign since they sum to zero). Consequently when $\BGj_0=-p\BI$
the maximum over $\BGz$ in \eq{5.17aa} is achieved when $\BA=0$ and taking the maximum over $\Ga$ gives
\beq -p\Tr\BGe_0 \geq p^2\left[\frac{1}{\Gk_1}+\frac{f_2}{\frac{\Gk_1\Gk_2}{\Gk_1-\Gk_2}+\frac{4f_1\Gm_1\Gk_1}{3\Gk_1+4\Gm_1}}\right],
\eeq{5.17ad}
or equivalently
\beq -p/(\Tr\BGe_0)\leq \Gk_{HSH}^+\equiv\Gk_1-\frac{f_2}{1/(\Gk_1-\Gk_2)-f_1/(\Gk_1+4\Gm_1/3)}, \eeq{5.17ae}
where $\Gk_{HSH}^+$ is the upper bulk modulus bound of \citeAPY{Hashin:1963:VAT} and
\citeAPY{Hill:1963:EPR}. The inequality \eq{5.17ad} can be 
rewritten as
\beq \frac{f_2}{-(\Tr\BGe_0/p)-1/\Gk_1}\leq \frac{\Gk_1\Gk_2}{\Gk_1-\Gk_2}+\frac{4f_1\Gm_1\Gk_1}{3\Gk_1+4\Gm_1},
\eeq{5.17af}
where we have used the fact that $-(\Tr\BGe_0/p)-1/\Gk_1$ is positive (since $\BGj_0:(\BCC^N)^{-1}\BGj_0\geq \BGj_0:(\BCC_1)^{-1}\BGj_0$
by \eq{5.8}). This then yields the volume fraction bound
\beq f_2\leq \frac{\frac{\Gk_1\Gk_2}{\Gk_1-\Gk_2}+\frac{4\Gm_1\Gk_1}{3\Gk_1+4\Gm_1}}{\frac{1}{-(\Tr\BGe_0/p)-1/\Gk_1}+\frac{4\Gm_1\Gk_1}{3\Gk_1+4\Gm_1}},
\eeq{5.18ag}
which we expect to be closest to the actual volume fraction when phase 2 (the softer phase) 
is the inclusion phase. Thus the
bound may be particularly effective for estimating the volume of cavities in a body. Note that
if some granules of phase 1 lie within these cavities, then such granules will not
contribute to this volume fraction estimate, but will contribute to the overall weight.
If the weight of the body has been measured (and the density of phase 1 is known)
this provides a way of estimating the volume of granules of phase 1 which lie within the
cavities. 

When multiple experiments have been
done, and the full tensor $\BCC^D$ or $\BCC^N$ has been determined, then the ``trace bounds'' of
Zhikov(\citeyearNP{Zhikov:1988:ETA}, \citeyearNP{Zhikov:1991:EHM}) and \citeAPY{Milton:1988:VBE} can be used. 
(These generalize the well known Hashin-Shtrikman (\citeyearNP{Hashin:1963:VAT})
bounds to anisotropic elastic composites.) Define the two traces
\beq \Tr_h\BCA=A_{iijj}/d,\quad\quad Tr_s\BCA=A_{ijij}-(A_{iijj}/d),
\eeq{5.17a}
for any fourth order tensor $\BCA$ with elements $A_{ijk\ell}$ in spatial dimension $d$.
Then, assuming the moduli of the two-phases are well ordered satisfying \eq{5.13a},
their lower and upper ``bulk modulus type bounds''
imply, through \eq{5.10}, the universal bounds
\beqa f_1\Tr_h[(\BCC^D-\BCC_2)^{-1}]
& \leq & \frac{1}{d(\Gk_1-\Gk_2)}+\frac{f_2}{d\Gk_2+2(d-1)\Gm_2}, \nonum
f_2\Tr_h[(\BCC_1-\BCC^N)^{-1}]
& \leq & \frac{1}{d(\Gk_1- \Gk_2)}-\frac{f_1}{d\Gk_1+2(d-1)\Gm_1},
\eeqa{5-18}
while their lower and upper ``shear modulus type bounds,'' imply the universal bounds
\beqa  f_1\Tr_s[(\BCC^D-\BCC_2)^{-1}] & \leq &
\frac{(d-1)(d+2)}{4(\Gm_1-\Gm_2)}+\frac{d(d-1)(\Gk_2+2\Gm_2)f_2}{2\Gm_2(d\Gk_2+2(d-1)\Gm_2)},\nonum
f_2\Tr_s[(\BCC_1-\BCC^N)^{-1}] & \leq &
\frac{(d-1)(d+2)}{4(\Gm_1-\Gm_2)}-\frac{d(d-1)(\Gk_1+2\Gm_1)f_1}{2\Gm_1(d\Gk_1+2(d-1)\Gm_1)}.
\eeqa{5.19}
Since these inequalities depend linearly on $f_1=1-f_2$ they can easily be inverted to obtain
bounds on $f_1$ given $\BCC^D$ or $\BCC^N$.

As noted by \citeAPY{Milton:1988:VBE} the lower and upper ``bulk modulus type bounds'' are tighter than those obtained by \citeAPY{Kantor:1984:IRB} and
\citeAPY{Francfort:1986:HOB}, which imply
\beqa \Tr_h\BCC^N&\leq& d\Gk_1-\frac{f_2}{\frac{1}{d(\Gk_1-\Gk_2)}-\frac{f_1}{d\Gk_1+2(d-1)\Gm_1}}, \nonum
1/\Tr_h[(\BCC^D)^{-1}]&\geq& d\Gk_2+\frac{f_1}{\frac{1}{d(\Gk_1-\Gk_2)}+\frac{f_2}{d\Gk_2+2(d-1)\Gm_2}}.
\eeqa{5.20}
For bodies $\GO$ of ellipsoidal or parallelopipedic shape the universal bounds \eq{5.20} were 
obtained by Nemat-Nasser and Hori (\citeyearNP{Nemat-Nasser:1993:MOP}, \citeyearNP{Nemat-Nasser:1995:UBO}):
see the equations (4.3.9) and (4.4.8), with $I=1$, in their 1995 paper. Their other bounds, with $I=2$,
which incorporate the ``shear responses'' of the tensors $\BCC^N$ and $\BCC^D$ are improved upon by the bounds \eq{5.19}
as can be seen using the inequality 
\beq \Tr_s\BCA^{-1}\geq (d-1)^2(d+2)^2/(4\Tr_s\BCA), \eeq{5.21}
which holds for any positive definite fourth-order tensor $\BCA$. 

\section*{Acknowledgements}
Eunjoo Kim is deeply thanked for generously providing figures \ref{3}, \ref{4}, and \ref{5}, and for doing the numerical 
simulations which generated them. Additionally the author is grateful to Hyeonbae Kang and 
Michael Vogelius for stimulating
his interest in this problem, and for their comments on an initial draft of the manuscript. 
The author is most thankful for support from the Mathematical Sciences Research Institute and the Simons foundation, 
through an Eisenbud fellowship, and from
National Science Foundation through grant DMS-0707978. 

\bibliography{/u/ma/milton/tcbook,/u/ma/milton/newref}

\end{document}